# YOURPRIVACYPROTECTOR: A RECOMMENDER SYSTEM FOR PRIVACY SETTINGS IN SOCIAL NETWORKS

Kambiz Ghazinour[1,2], Stan Matwin[1,2,4] and Marina Sokolova[1,2,3]

[1] School of Electrical Engineering and Computer Science University of Ottawa
[2] CHEO Research Institute, Ottawa, Ontario, Canada
[3] Faculty of Medicine, University of Ottawa, Ontario, Canada
[4] Faculty of Computer Science, Dalhousie University, Halifax, Nova Scotia, Canada

`kghazino@uottawa.ca` , `stan@cs.dal.ca` , `msokolova@ehealthinformation.ca`

## ABSTRACT

*Ensuring privacy of users of social networks is probably an unsolvable conundrum. At the same time, an informed use of the existing privacy options by the social network participants may alleviate - or even prevent - some of the more drastic privacy-averse incidents. Unfortunately, recent surveys show that an average user is either not aware of these options or does not use them, probably due to their perceived complexity. It is therefore reasonable to believe that tools assisting users with two tasks: 1) understanding their social net behaviour in terms of their privacy settings and broad privacy categories, and 2) recommending reasonable privacy options, will be a valuable tool for everyday privacy practice in a social network context. This paper presents YourPrivacyProtector, a recommender system that shows how simple machine learning techniques may provide useful assistance in these two tasks to Facebook users. We support our claim with empirical results of application of YourPrivacyProtector to two groups of Facebook users.*

## 1. INTRODUCTION

### 1.1 Privacy on social networks

Modern social network and services have become an increasingly important part of how users spend their time in the online world. The social network is a proper vehicle for people to share their interests, thoughts, pictures, etc. with their friends or the public. While sharing information about the self is intrinsically rewarding [12], the risk of privacy violation increases due to disclosing personal information [5,13]. Recent cases, such as Canada's Privacy Commissioner challenge to Facebook's privacy policies and settings , have shown a growing interest on the part of the public with respect to how social network and services treat data entrusted to them. Some of the privacy violation incidents could be mitigated or avoided if people used more privacy setting options [8].

Facebook with current number of 955 million users and still growing is the most popular social network and as such motivates our work on privacy settings and issues. Over the past several years, Facebook has provided many privacy settings and options for the users, e.g. users can determine who can see their personal information such as their date of birth or home town; they can also set to whom their photo albums can be visible to. Unfortunately most users do not know the importance of privacy settings, do not have enough time to read and comprehend tedious and long pages of privacy settings or simply do not understand how these settings available for them work [8]. It also becomes more concerning when we realize that the default privacy settings for the posts, photo albums, etc. are set as being visible to the public.

## 1.2 Facebook

In May 2012, Consumer Reports Magazine[1] surveyed online 2,002 US households, including 1,340 that are active on Facebook. The survey results were extrapolated to estimate national totals and given in terms of absolute numbers with respect to the U.S. population (169 million monthly active users of Facebook in the U.S. as of March 31, 2012). The privacy protection results raise some concerns as follows:

1) Some people are sharing too much. 4.8 million users have used Facebook to say where they planned to go on a certain day which is a potential tip-off for burglars; 4.7 million *liked*[2] a Facebook page about health conditions or treatments (details an insurer might use against them).

2) Some people do not use privacy controls. Almost 13 million users said they had never set, or did not know about, Facebook's privacy tools. And 28% shared all, or almost all, of their wall posts with an audience wider than just their friends.

3) And problems are on the rise. 11% of households using Facebook said they had a privacy-related trouble during the last year (2011), ranging from someone using their log-in without permission to being harassed or threatened. That projects to 7 million households -30% more than the previous year (2010).

Although these results were inferred based on the data collected from the users in the United States, nothing suggests the results for the rest of the world would be less concerning.
Our approach to remedy this situation is to develop a tool that monitors and suggests a privacy setting to the user rather than leaving the privacy settings as default or even setting them too loose that basically little privacy, if any, is protected.

## 1.3 YourPrivacyProtector

In this paper, we present a recommender system for privacy setting that suggests privacy settings that have been automatically learned for a given profile (cluster) of users.

Our tool, called YourPrivacyProtector, uses monitoring of the privacy settings and a consequent machine learning of the user profiles to recommend an optimal setting for a particular user. In other words:

YourPrivacyProtector allows users to see their current privacy settings on their social network profile, namely Facebook, and monitors and detects the possible privacy risks. It monitors by providing a brief review for the users (in their user interface).

The tool acts as a recommender system. It shows to a user the attributes that play important role in setting privacy preferences for individuals on a social network. Recommending a privacy setting occurs based on the notion of collaborative filtering and the similarity of the preferences chosen by the user who desires to set the privacy setting and the other users who share common preferences. Preliminary discussion of this work had been reported in [4].

---

[1] http://www.consumerreports.org
[2] When you *like* a Facebook page it means you are interested in that

The rest of this paper is organized as follows. Section 2, describes some related work in this area. Section 3, introduces data privacy. Section 4 discusses our approach in the profiling phase and classification phase. Section 5 describes our empirical study and the results from our model. Section 6 describes how the recommender system works and finally, Section 7 concludes the work and discusses some future research directions.

## 2. RELATED WORK

Recommender systems for privacy settings have started to attract researchers' attention in recent years. As more options are given to the users to set their privacy preferences, users are more confused, frustrated or sometimes simply ignorant about setting them. As in the Facebook case, the privacy settings are hard to be set and users with average knowledge about computers cannot easily find or set the privacy settings as they should be.

In [2], the authors introduce the privacy policy simplification problem and presented enList, a system that uses automatically extracted friend lists to concisely represent social network privacy policies. They also conducted a laboratory-based user study to evaluate the effectiveness of the concise representation compared to a verbose representation. Their study demonstrated that their method resulted in better accuracy for policy comprehension, recollection and modification tasks.
In [9], the authors present a dynamic trust-based privacy assignment system which assist people select the privacy preference on-the-fly to the piece of content they are sharing, where trust information is derived from social network structure and user interactions. Their model using a cosine similarity function detects a two-level topic sensitive community hierarchy and then assigns privacy preference for users based on their personalized trust networks. They demonstrate the feasibility and effectiveness of their model on a social object network dataset collected from Flickr.

In another example, [10] propose an intelligent semantics-based privacy configuration system, named SPAC, to automatically recommend privacy settings for social network users. SPAC learns users' privacy configuration patterns and make predictions by utilizing machine learning techniques on users' profiles and privacy setting history.

Recently [3] study the problem of how smartly sharing information in online social networks without leaking them to unwanted targets. They formulate this problem as the optimization problem, namely maximizing a circle of trust (MCT), of which they construct a circle of trust to maximize the expected visible friends such that the probability of information leakage is reduced to some degree.

## 3. USER PRIVACY ON THE NET

### 3.1. Internet privacy and user preferences

Ideally, a definition for internet privacy for users would be the ability to control (1) what information one reveals about oneself and (2) who can access that information. Essentially, when the user's data is collected or analyzed without the knowledge or consent of, the user, her/his privacy is violated. When it comes to the usage of the data, the user should be informed about the purposes and intentions for which the data is being or will be used. The last but not the least: when a data collector wants to disclose the data to other individuals or organizations, it should be done with the knowledge and consent of the user.

Personal information disclosure and the internet privacy issues become even more obvious in online social networks. In [5], analyzing the content of 11,800 MySpace posts, has shown that many users extensively reveal personal health information. [6] analyzes the online behavior of more than 4000 Carnegie Mellon University students who are member of Facebook. The authors evaluate the amount of information the students disclose and study their usage of the site's privacy settings. Their study reveals that a large number of the participants are unconcerned or simple pragmatic about their privacy.

People have different privacy concerns. Therefore, there is no a single privacy policy that fits every user. For instance, one user may be concerned about revealing the home phone number to a potential third party, whereas another one may not be concerned. Alan Westin, known for his works in privacy and freedom, has conducted over 30 privacy surveys [7]. In his work, Westin has classified people into three groups: High and Fundamentalist, Medium and Pragmatist, Low and Unconcerned. Privacy fundamentalists are described as unwilling to provide any data on web sites and are highly cautious about revealing their personal information. Privacy pragmatists are willing to compromise their privacy if they gain some benefits in return. And the third group consists of people are those unconcerned with their privacy at all and are willing to reveal and disclose any information upon request. These surveys demonstrate that more people are getting concerned about their privacy because they feel they are losing control over their data.

### 3.2. Privacy dimensions

We review the key dimensions of the user data privacy. This helps us to identify what elements should be involved in measuring a privacy preference. [1] introduces a data privacy taxonomy that captures *purpose*, *visibility* and *granularity* as the main privacy elements.

- Purpose defines the intention of the data provider for how data can be used after collection (e.g., members provide their mailing address to Amazon.com for the purpose of "shipping orders").
- Visibility defines who is allowed to see the provided data (e.g., members of Facebook can specify what group of people can visit their profile, friends, friends of friends and etc.). Visibility of data is an important key in ensuring appropriate system utility.
- Granularity of data defines how much precision is provided in response to a query (e.g., data providers could define whether their exact age is shown or a range such as child, teenager, adult).

Currently Facebook and other social networks provide users with the privacy settings that help to set their privacy preferences. The privacy settings, which we differentiate from the privacy policies, define mostly visibility and in rare cases granularity of the data. For instance, in Facebook the users can choose that only the day and month (and not year) of their date of birth will be visible to their friends. This prominence of visibility suggests that we focus on the visibility feature of data. We consider the way the visibility features has been set by a user to be an indicator of how important the privacy is for that individual.

## 4. UTILIZING USER PROFILES

### 4.1. Building user data

In building a set of user data, we were interested in the following three types of the user parameters:

International Journal of Security, Privacy and Trust Management ( IJSPTM) Vol 2, No 4, August 2013- *User's personal profile*; There are several attributes stored in a user profile, ranging from user's ID and name to the work experience and even the time zone of where the user resides. Table 1 shows the attributes we collected from Facebook, i.e. personal profiles, their formats and pre-defined values.
- *User's interests*; on each Facebook profile users have the option of expressing what they are interested in. For instance, movies they like to watch, books that they read, music they listen to, sport they do or watch, and many other activities. We collected the interest id and its category that it belongs to. There are around 196 distinct categories defined in Facebook. Since each item has a unique ID it can be compared among people to realize who shares the same interest(s). Table 2 shows the attributes we collect from Facebook regarding the user's interests and their descriptions.
- *User's privacy settings on photo albums*; The Album object has several attributes including the title, description, location, cover photo, number of photos, created time, and etc. We only collected the name and privacy settings of the album which had predefined values as shown in Table 3.

These parameters will help us to construct the personal profile of a user in Section 4.2. Later on the recommender system will utilize the collected data to find the similarities between the individuals. Note that we also wanted to build the user data set using User's privacy settings on posts. However, Facebook application programming interfaces (APIs) are very restricted in what they allow to scan from the posts. Often API only reveals what the post stored in a recent time, i.e., it returned zero posts if the participant posted no comments or links in the past 24 hours. Hence, we were not able to add the privacy settings on posts to the user profile.

**Table 1** – Attributes collected from user's profile

| Attribute | Description |
|---|---|
| Name | The user's full name. `string`. |
| Gender | The user's gender: 'female' or 'male'. 'string'. |
| Birthday | The user's birthday. `user_birthday` or `friends_birthday`. Date `string` in `MM/DD/YYYY` |
| Education | A list of the user's education history. `user_education_history` or `friends_education_history`. `array` of objects containing `year` and `type` fields, and `school` object (`name`, `id`, `type`, and optional `year`, `degree`, `concentration` array, `classes` array, and `with` array ). |
| Hometown | The user's hometown. `user_hometown` or `friends_hometown`. object containing `name` and `id`. |
| Location | The user's current city. `user_location` or `friends_location`. object containing `name` and `id`. |
| Political | The user's political view. `user_religion_politics` or `friends_religion_politics`. `string`. |
| Relationship | The user's relationship status: `Single`, `In a relationship`, `Engaged`, `Married`, `It's complicated`, `In an open relationship`, `Widowed`, `Separated`, `Divorced`, `In a civil union`, `In a domestic partnership`. `user_relationships` or `friends_relationships`. `string`. |
| Religion | The user's religion. `user_religion_politics` or `friends_religion_politics`. `string`. |

**Table 2 -** Attributes collected from users interests

| Attribute | Description |
|---|---|
| Interest ID | ID of the interest. 'string' |
| Interest Category | Name of the category that the interest belongs to such as artist, car, restaurant, movie, etc. 'string' |

**Table 3 –** Attributes collected from photo albums

| Attribute | Description |
|---|---|
| Name | Title of the album. 'string' |
| Privacy value | EVERYONE, FRIENDS, FRIENDS_OF_FRIENDS, NETWORKS_FRIENDS, CUSTOM. 'string' |

### 4.2. Representation of a user

For each user *u*, we represent the profile as a vector *Vu* as follows:
*Vu = [Du , PRu]*  where *Du = [Pu , Iu]*.

*Du* consists of the set of Pu (user's profile attributes) and *Iu* (the set of user's interests). *PRu* is the set of user's privacy setting for their photo albums. For instance:

*Pu= [gender, age, location, relationship, education, political view];*
*Iu = [interest categories: interest instances];*

*PRu=[number of photo albums, number of albums visible to friends of friends, number of albums visible to public].*

To showcase our data, we randomly selected three user profiles that we collected for our empirical experiment (see Section 5). In Example 1, to protect the privacy of the individuals we change their real name.

**Example 1 –** The vector V for three users, Sandrine, John, and Alice might appear as follows:
$V_{Sandrine}$=[Female, 30, Canada, -, Grad, Liberal, TV: The Big Bang Theory, Sport: Basketball - Golf; 20,4,5].

$V_{John}$ =[Male, 30, Canada, Single, Grad, Liberal, TV: The Big Bang Theory, Sport: Basketball - Golf; 25,1,2].

$V_{Alice}$ =[Female, 38, Canada, Single, Grad, -, TV: The Big Bang Theory, Sport: Basketball - Golf; 2,0,1].

### 4.3. Challenges of the user profiles

Some attributes have not been defined by the users in their Facebook profiles. This does not mean that they have not allowed our application to see the value; they simply have not entered any value for that. Although such omission makes the job of the similarity function on the training set harder, the absence of certain attribute values relays a message about the user attitude towards the attribute. We utilize the absence of values when *YourPrivacyProtector* suggests a better privacy setting for an individual whose profile is examined against our set.

From the Example 1, Sandrine has not entered any value for her relationship status, which clearly demonstrates that she does not feel comfortable sharing that which is a privacy protection gesture. When nothing is shared, no privacy violation occurs too!

Another important aspect of finding similarity between the individuals' profiles is the importance and weight of the attributes. For instance, does gender play a bigger role or age? Are Alice and Sandrine as two ladies more similar or John and Sandrine as two 30 year old individuals?
To answer these questions we will check the similarity of the privacy settings which will be tested from two different perspectives:

- *existence of the values* (e.g. if the photo album exist)
- *value closeness* (is it visible to the public or friends of friends?)

We measure privacy preferences of the users by examining the privacy settings they have chosen for their photo albums. We count the total number of albums and check to see if there any of them are set visible to public, friends of friends, friends or simply customized otherwise and also calculates the ratio. For instance, assume John has 25 photo albums and only one of them is set visible to the public and Alice has 2 albums from which one is set visible to the public. We can predict that probably John is more cautious about his privacy of the albums than Alice. Furthermore, if Sandrine has put no photo albums this might indicate that she is even more cautious than John. The non-existence of the values can be also interpreted as a privacy behavior. Since posting photo albums on Facebook is a very common act and very easy to be done the assumption that Sandrine did not know how to put photo albums on her profile to justify why she has 0 photo albums does not seem to be a valid reason.

We also emphasize that in our approach the missing values (i.e., the data that the user did not provide on their profile ($Pu$)) can play as an important role as the posted values. For instance, the relationship status and political view were not provided by Sandrine and Alice respectively. Although this is partially captured by the difference in the data representation and in resulting similarity functions, these missing values can be used in recommendation to an individual who seeks advice from the system. For instance, if John is more similar to Sandrine in terms of the user profile, then *YourPrivacyProtector* not only recommends that John should have a privacy setting similar to Sandrine for the photo albums, but also advices John not to reveal his relationship status. This is the same pattern that Sandrine followed.

### 4.4. Classification of the user profiles

At the preliminary step, we screen the users based on their attitude toward privacy. As discussed in Section 3.1, Westin categorizes people into three different groups of fundamentalists, pragmatics and unconcerned. In the current work we keep the same taxonomy and examine the users' attitude towards sharing their photo albums as an indication of their privacy values and to which privacy category they more likely belong to. We use this privacy category since it is broad enough to address all the privacy attitudes and also it has been well-studied and well-addressed in the data privacy community. We also use photo albums since users treat them as a very personal and tangible type of personal identifiable information. Furthermore, it is one of the data items that Facebook allows us to check its privacy settings using the Facebook API functions.

To perform the profiling phase, we look at each individual's ratio of albums visible to public, friends of friends, friends, and custom. For simplicity we use the following rule to determine to which of the three privacy categories the user belongs to:

```
If # of photos shared == 0 then:
        The user's privacy_category = Fundamentalist.
```

Else if ratio of photos visible to friends + ratio of photos visible to custom > %50 then:
    The user's privacy_category = Pragmatic.
Else:
    The user's privacy_category = Unconcerned.

After gathering values for the privacy_category attribute of each user, we use the Decision Tree (DT) to infer the profile type of each user. For the recommendation task, we apply the k-nearest neighbor algorithm (KNN). The idea is that when a user joins the Facebook we put their profiles in the KNN classifier and determine to which privacy setting class they belong to. Then, considering the specifications of that class, our recommender system suggests the user what data could be disclosed and which ones should not be shared.

## 5. EMPIRICAL DATA

### 5.1. Data harvesting

In our empirical study, the first dilemma was to collect information and create a training set. We implemented a Facebook application written in JavaScript and PHP in order to access the Facebook user`s profile and settings. In 2010, Facebook introduced the concept of Open Graph which allows apps to model user activities based on actions and objects. The JavaScript SDK provides a rich set of client-side functionality for accessing Facebook's server-side API calls. These include all of the features of the Graph API.

Using the JavaScript SDK we had to initialize a connection between our app and the Facebook server using the FB.init() and then authenticate and authorize our application with the application ID and other information that were previously provided using the App Dashboard. Next, using the FB.api() we used Graph APIs query to obtain information about the Facebook users. It should be noted that besides using Graph APIs one can use the Facebook Query Language (FQL) with a SQL-style interface to query the data exposed by the Graph API. Although it provides some advanced features not available in the Graph API, including batching multiple queries into a single call, we preferred to use the Graph APIs for simplicity purposes.

### 5.2. Participating users

We asked 150 undergraduate students from three universities (two in Canada and one in Brazil) via email to participate in this research if they have a Facebook profile and we informed them that their privacy is preserved and the data will be anonymized in case it is used in a publication.

We informed them that the overall goal of this research is to develop a recommender system that assists people in using privacy settings in popular social networks, specifically in Facebook. The participants were told that in this project we wanted to investigate if there is a relationship between the personal information and interests of the Facebook users, and the way they choose their privacy settings. They were informed that this research requires some basic data collected on a voluntary basis from Facebook users. We explained that as the project was done in a university environment, we naturally focused on students and that was why we asked their help with data collection. They were also informed about the following procedures:

- They do not need to answer any questionnaire or survey. After they log in to their Facebook account and enter the address https://apps.facebook.com/privacy_check/ they just need to give their consent for this application to access to their data and privacy settings.

- They needed to run this application only once – since the data collection phase happened just for one time. The application extracted some basic information that discussed in Section 5.1 and the information was stored in a comma separated text file on the server which was password protected.

In the next phase, our data harvesting application went through the photo albums of the participant and retrieved the number of albums and the privacy setting of each album.

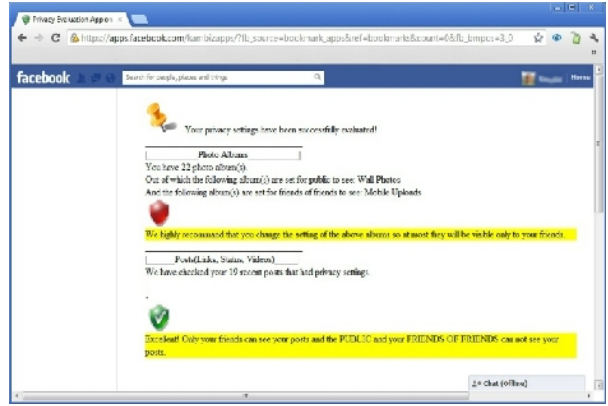

**Figure 1 –** The application interface

The application output was a brief report to the participant, listing the photo albums which were visible to everyone (public) or friends of friends. We assumed that when the user created a photo album and uploaded photos, they intended their friends (or friends in their network) to see them. If the privacy setting was set to 'friends of friends' or 'everyone', then the report notified the user to tighten privacy settings by referring the user to the name of the photo album. Figure 1 illustrates a snapshot of the report interface. In this example the application found two problems in the photo albums. The album "wall photos" are set visible to the public and the album "mobile uploads" are set visible to friends of friends. The report recommends that the user tightens the privacy setting of these two albums to reduce potential privacy breaches.

The participants were also told that the collected data will not be disclosed to any third-party and will be anonymized to be used for research purposes only.

## 5.3. Data analysis

Before running the classification algorithm we need to perform some pre-processing tasks to make the data ready for classification. In the data collection phase, the Facebook members mainly speak Portuguese or English and for some fields such as religion, they had same values in different languages (e.g. Christian-Catholic vs Cristão-Católico) that we had to fix manually. In the field education, the users occasionally entered several education levels ranging from high school to graduate studies in no particular order.  We were interested in the highest education level so we had to read the whole field and select the highest one as the education of that Facebook member.

Another task to be performed was to sort the interest categories and the instances the user chose in each category. That analysis further will help the classification algorithm to classify the users based on their interest as well. The collected data had some unique features with respect to the attributes that users felt comfortable sharing. As illustrated in Table 4, among all the attributes we collected, age, i.e. year of birth, had 0% missing value.

**Table 4 –** % of the values users did not share

| Attribute | % missing |
|---|---|
| Degree | 94% |
| Political view | 91% |
| Religion | 67% |
| Relationship | 39% |
| Hometown | 36% |
| Interests | 32% |
| Location | 20% |
| Education | 12% |
| Gender | 4% |
| Age | 0% |

This means all the users provide their year of birth. The degree attribute has the most missing values, namely 92% of users do not specify their degrees. It is an interesting fact because only 12% of the users have not provided their education level. Users are also reluctant to share their political view, religion and relationship status. It was not surprising to see people are less concerned with disclosing their gender, education and location. However, the fact that only 28% did not reveal their interests indicates that this can be used as important criteria for our classification algorithm. We found the top 25 most common interests among the users; Table 5 reports the top 10. The interests are mainly in categories such as TV shows, musician band, and sports. We also performed a classification based on these common interests to observe if there is a mapping between the interests and the way users chose privacy setting for the photos albums. We are aware that interests are subject to change and highly depend on the current trend (e.g. a music or video clip that goes viral, a TV show that becomes popular, etc.). The fact that a person who liked a particular music or TV show that is not popular anymore still reveals some information about the taste of that individual and may results in predicting other interests which are currently popular but have not selected by him/her yet.

**Table 5 –** Top 10 topics that users show interest

| # of users interested | Topic | Category |
|---|---|---|
| 10 | The Big Bang Theory | TV show |
| 7 | Game of Thrones | TV show |
| 6 | Dexter | TV show |
| 5 | Coldplay | Musician band |
| 5 | FRIENDS | TV show |
| 5 | Suave Sabor | Restaurant café |
| 5 | How I Met Your Mother | TV show |
| 5 | Adele | Musician band |
| 4 | Bob Marley | Musician band |
| 4 | Chico Buarque | Musician band |

Figure 2a shows the ratio of photos that are set visible to the public and friends of friends (FOF) which are considered weak privacy settings and may lead to privacy breach. Remarkably, the number of photo albums that are set visible to the public, marked by the red line in graph, is

20

considerably higher than the number of photo albums that are set visible to friends of friends. One possible explanation for this phenomenon is that the default privacy setting for photo albums on Facebook is set to 'public'. Since most users did not bother or care to change the default setting, the number of photo albums that are visible to public is higher.

Figure 2b shows the ratio of photos that are set visible to only friends or customized for a particular list of friends which are considered a stronger privacy setting. We observed that the numbers of photo albums that are set visible to friends or have been customized to be visible to a particular group of friends are fairly balanced and there is no predominant group detected.

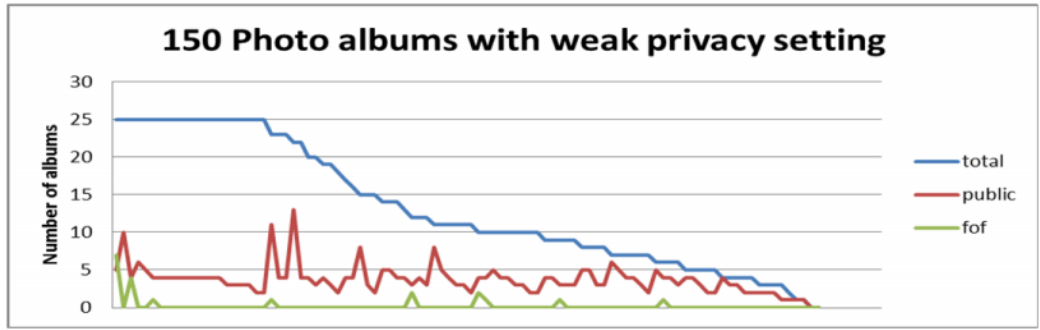

a) Visible to public or friends of friends

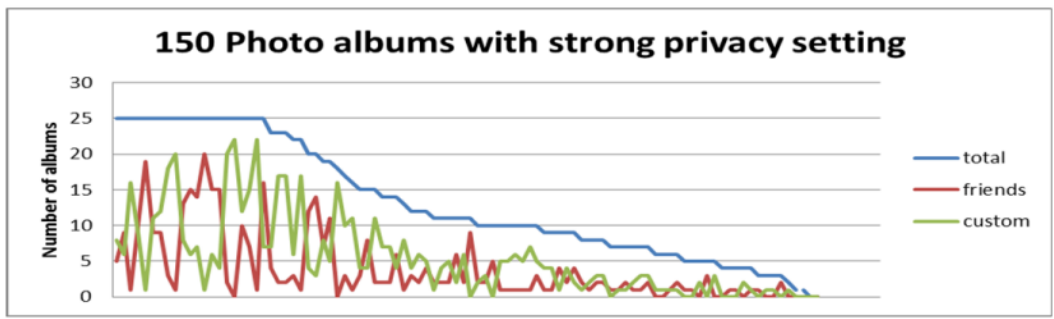

b) Customized or set visible to just friends

**Figure 2-** Distribution of the number of photo albums and their privacy settings.

## 6. EMPIRICAL EVALUATION OF *YOURPRIVACYPROTECTOR*

### 6.1. User classification

With the data collected from 150 users, as described in Section 5, we can train our classifier to categorize a new Facebook member based on his/her personal profile data. We used Decision Trees to categorize him/her into the three types of privacy behaviors: ignorant, pragmatic and fundamentalist. The resulting Decision Tree provides us with profiles for these privacy categories. Such profiles help the understanding of attitudes towards privacy in terms of user's personal characteristics and interests.

Figure 3 shows such a tree. Based on the attributes provided, age, gender and the interests to two topics (Adele, the singer, and Friends, the TV show) are learned to be important criteria. The resulting tree shows

- if the participant has less than 21 years old then it is most likely that he/she belongs to privacy ignorant group

- if the individual is older than 21:
o if she is female it is more likely to have a privacy pragmatic attitude
o if he is male then his privacy behavior depends on his favorite singer (e.g. Adel) or his favorite TV show (e.g. Friends) .

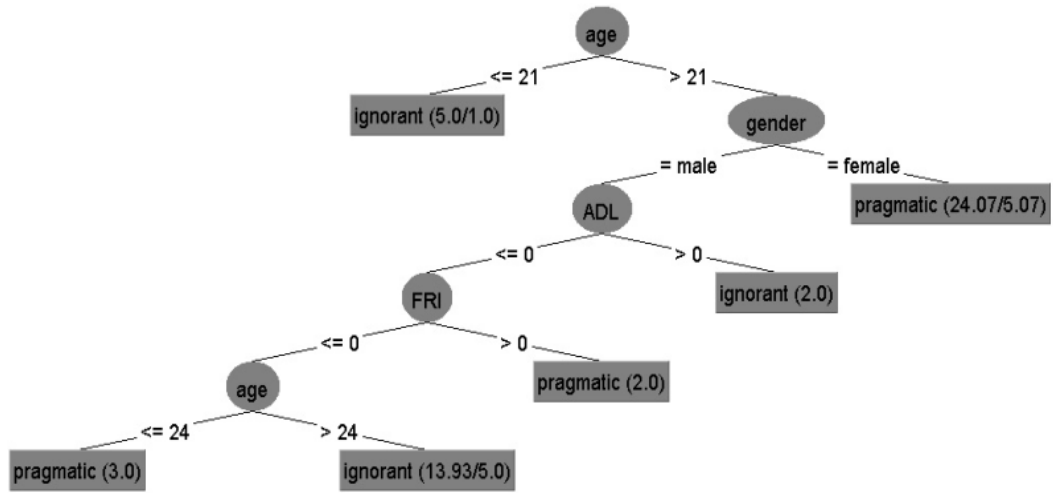

**Figure 3** – A sample decision tree for analyzing criteria of privacy attitudes.

We realize that the relation between categories of user interests and their privacy behavior can be more complex than has been shown here. Further studies and analysis of more data can help to explore such relation that happens especially in social media and social networks.

## 6.2. Recommending privacy settings

As another benefit of *YourPrivacyProtector*, we have used a simple approach to recommend a binary privacy setting based on the use of the k-nearest neighbor (KNN) classifier to suggest that value. Note that the use of KNN makes this classifier work like a collaborative filtering recommender system [11].

We used the KNN in Weka with K=3 to classify the participants based on their common personal information they have disclosed, such as age, education, hometown and their preferences such as political views, religion and etc. Hence, our training data set would be the individuals with binary values of their profile attributes either disclosed or not. Due to our small set data, we performed five replications of twofold cross-validation. In each replication, our collected data set were randomly partitioned into two equal-sized sets in which one was the training set which was tested on the other set. Then we calculated the average and variance of those five iterations for three different attributes of education, location and relationship status. Table 6 shows the results of this 5X2-fold cross validation.

**Table 6** – 5X2-fold cross validation

| Attribute<br>Average 5X2 | Education | Location | Relationship status |
| --- | --- | --- | --- |
| Correctly classified instances | 88% | 84% | 62% |
| Mean absolute error | 0.1772 | 0.2488 | 0.4274 |

The following example clarifies the way the recommender system suggests user's privacy settings to disclose an attribute or not.

**Example2–** Alice (from Example 1) has entered her information on her Facebook profile. Her profile shows that she has disclosed the following attributes: gender, age, hometown, relationship status, education and some topics of interest. *YourPrivacyProtector* also recognizes that Alice has not disclosed her political views, and religion. The model puts the information through the 3-nn classifier in the pool of 150 Facebook profiles. The system finds the three nearest neighbors whose profile is similar to Alice's:

| Gender | Age | Hometown | Relationship | Politics | Education | Interests |
|--------|-----|----------|--------------|----------|-----------|-----------|
| Female | 38 | Canada | Single | - | - | TV: The Big Bang Theory, Sport: Basketball; |
| Female | 30 | Canada | Single | - | - | TV: The Big Bang Theory, Sport: Golf; |
| Female | 35 | Canada | Single | - | - | TV: The Big Bang Theory, Sport: Basketball Golf; |

Based on the information from the three nearest neighboring members, *YourPrivacyProtector* recommends that Alice should not disclose her education attribute either.

In brief, based on the information she has disclosed and the topics that she has shown interest in, *YourPrivacyProtector* uses a 3-nearest neighborhood classifier and finds the three closest profiles in which they have disclosed same attributes. Knowing those three records, the system identifies if those profiles disclose education level or not and recommends Alice to do so.

In case of conflicting settings for disclosure of an attribute (i.e. some neighboring members disclose and some do not), different actions could be taken to resolve it. For instance, if the majority of the neighboring items do not disclose, the systems recommends not disclosing the data or for example even if there exists one member who does not disclose the attribute then the system recommends a tighter privacy attitude and basically suggest that the member not to disclose the attribute.

It should be pointed out that our recommender system is only interested to tighten the privacy setting and for example never recommends that an attribute that the user has decided not to share, should be disclosed.

## 7. CONCLUSIONS AND FUTURE WORK

In this research work, we have introduced *YourPrivacyProtection* that works as a monitoring and recommender system. We have applied it to recommend privacy settings in a social network Facebook. Since a large portion of Facebook users do not change the default privacy settings on Facebook, our system's recommendations can help users to improve the privacy settings rather than leaving the privacy settings "as is" ,which is set visible to public, or even setting them so loose that little privacy, if any, is protected.

Next, we performed an empirical study on a sample data to demonstrate the feasibility of our model, proof of concept and empirically evaluate the system. We asked 150 undergraduate students from three universities to run our Facebook app on their profile. Our study showed that participants were more willing to share information about their age, gender and education and less willing to disclose their religion, political view and degree. Furthermore, we used Decision Tree to show that how people are classified based on their privacy preferences which was derived from the way they shared their photo albums. We also demonstrated that how a KNN (where K=3) algorithm was used in the recommender system that suggests privacy settings for an individual Facebook member.

Our vision for the future deployment and use of the recommender part is as follows. In a given population (e.g. High-school students) a group of privacy-aware volunteers would make their profiles available as a training set for our system. Note that just a one-time read of the profile of the volunteer participants is involved in building the training set. Subsequently all other members of this population could use suggestions from the system for their settings of profile elements (i.e., disclose or do not disclose). As these suggestions would be based on informed decisions of more privacy-aware but otherwise similar members of the same population, they would appear reasonable to the users and still protect their privacy.

We plan future experiments with a larger number of users to demonstrate the scalability of our model. We also consider that some information details are more sensitive than others; for example, personal health information or exam grades are more sensitive that the photo albums for certain individuals. It will also be an important project to study how sensitive personal data is disclosed and its connection with the privacy behavior and privacy categories of the users.

As another future research direction we would like to perform a deeper analysis on the profiling phase of our study. We understand that existence of photo albums that are set visible to public by itself may not be a good indicator of a privacy unconcerned user. It is possible that the user takes pictures of the nature or his art works and set it visible to the public which does not imply any privacy violations. In the next step we would examine if the user or other individuals closely related to them are tagged or mentioned in the pictures that sharing them may result in privacy violations.

## AUTHORS

**Kambiz Ghazinour**

Dr. Ghazinour is a Post-doctoral Fellow in the area of Data Privacy at the Text Analysis and Machine Learning Research group, in School of Electrical Engineering and Computer Science, University of Ottawa, Canada. He is also member of The Electronic Health Information Laboratory, CHEO Research Institute in Ottawa. He Obtained his PhD from the Department of Computer Science, University of Calgary in 2012. Since 2008 he has done research in data privacy and published several papers in this field. He has been awarded many research scholarships. He is interested in privacy issues in social networks, electronic healthcare systems and big data analysis.

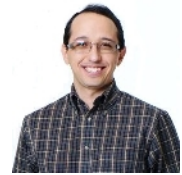

**Stan Matwin**

Dr. Matwin is a Canada Research Chair (Tier 1) at the Faculty of Computer Science, Dalhousie University, and the Director of the Institute for Big Data Analytics. He is also Emeritus Distinguished Professor of Computer Science at the University of Ottawa, and a Professor in the Institute for Computer Science of the Polish Academy of Sciences. His research is in machine learning, data mining, and their applications, as well as in technological aspects of Electronic Commerce. Author and co-author of over 200 research papers, he has worked at universities in Canada, the U.S., Europe and Latin America, where in 1997 he held the UNESCO Distinguished Chair in Science and Sustainable Development. Former president of the Canadian Society for the Computational Studies of Intelligence (CSCSI) and of the IFIP Working Group 12.2 (Machine Learning). ). Recipient of a CITO Champion of Innovation Award. Programme Committee Chair and Area Chair for a number of international conferences in AI and Machine Learning. Member of the Editorial Boards of the Machine Learning Journal, Computational Intelligence Journal, and the Intelligent Data Analysis Journal.

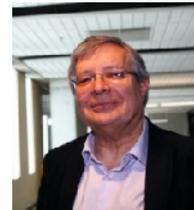

**Marina Sokolova**

Dr. Sokolova works in Text Data Mining and Machine Learning where she publishes extensively in areas related to text data mining of personal health information and information leak prevention. Dr. Sokolova obtained her PhD from the School of Information Technology and Engineering, University of Ottawa in 2006. She has been awarded with grants and merit-based awards from Natural Sciences and Engineering Research Council, Canadian Institutes of Health Research, and Japan Society for the Promotion of Science. She is a member of program committees of international conferences on Artificial Intelligence and reviews for international journals in the field of Text Data Mining. Starting in 2003, she published 5 book chapters, 9 journal papers and many referred conference papers.

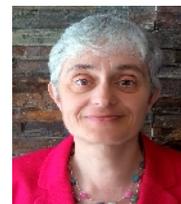